\documentclass[twocolumn,floatfix, showpacs]{revtex4}

\usepackage[final]{epsfig}
\usepackage{amsmath}
\usepackage{parskip}

\begin{document}

\title{Through the blackness --- high $p_T$ hadrons probing the central region of 200 AGeV Au-Au collisions}

\author{Thorsten Renk}
\email{trenk@phys.jyu.fi}
\affiliation{Department of Physics, PO Box 35 FIN-40014 University of Jyv\"askyl\"a, Finland}
\affiliation{Helsinki Institut of Physics, PO Box 64 FIN-00014, University of Helsinki, Finland}

\pacs{25.75.-q}

\begin{abstract}
The energy loss of high $p_T$ partons propagating through a hot and dense medium is regarded as a valuable tool to probe the medium created in ultrarelativistic heavy-ion collisions. The angular correlation pattern of hadrons associated with a hard trigger in the region of $p_T \sim$ 1-2 GeV which exhibits a dip in the expected position of the away side jet has given rise to the idea that energy is lost predominantly to propagating collective modes ('Mach cones'). Recent measurements by the STAR collaboration have shown that for a high $p_T >$ 8 GeV trigger the angular pattern of associate hadrons for $p_T > 4$ GeV shows the emergence of the expected away side peak. These di-jet events suggest that the away side parton may emerge occasionally without substantial energy loss. Since in such a back-to-back configuration one of the partons may travel through the central region of the fireball, the average in-medium pathlength is substantial and the expected energy loss is not only sensitive to the initial geometry of matter but also to the change of geometry due to expansion. We show that radiative energy loss is able to explain the dijet events provided that the expansion of the medium is taken into account.
\end{abstract}

\maketitle

\section{Introduction}
\label{sec_introduction}

Announcements have been made by all four detector collaborations at  RHIC \cite{RHIC-QGP} that a new state of matter, distinct from ordinary
hadronic matter has been created in ultrarelativistic heavy-ion collisions (URHIC). A new and exciting challenge for both
experiment and theory is now to study its properties. The energy loss of hard partons created in the first moments
of the collision has long been regarded a promising tool for this purpose \cite{Jet1,Jet2,Jet3,Jet4,Jet5,Jet6}. However,
before this tool can be fully exploited, the influence of the medium evolution on the energy loss has to be understood
in a more quantitative way (cf. \cite{Jet-Flow}).

Recently, measurements of two-particle correlations involving one hard trigger particle have shown a surprising
splitting of the away side peak for all centralities but peripheral collisions, qualitatively very different from
a broadened away side peak observed in p-p or d-Au collisions \cite{PHENIX-2pc}. Interpretations in terms of energy
lost to propagating colourless \cite{Mach,Shuryak, Stoecker} and coloured \cite{Wake} sound modes have been suggested 
for this phenomenon, and calculations within a dynamical model evolution have shown that the data can be reproduced under the assumption that a substantial amount of lost energy excites a sonic shockwave \cite{Mach}.

However, the STAR collaboration has reported \cite{Dijets, Dijets2}  that the picture of two-particle correlations is qualitatively changed if both transverse momentum $p_T$ of the trigger hadron and transverse momentum cut of associate hadrons are increased. Specifically, for $p_T^{\rm trigger} >$ 8 GeV and $p_T^{\rm assoc} > $ 4 GeV a clear signal at the expected position of the away side jet has been observed. This suggests that at such large momenta the away side parton is not always absorbed by the medium but may emerge with a substantial amount of energy left. Since radiative energy loss is based on a probabilistic picture of loosing the amount of energy $\Delta E$ given the parton path through the medium, we note that there is always a finite (but small) probability that a parton does not loose any energy even if it propagates through the densest central region.

For a realistic theoretical model of dijet events it is necessary to note that for a typical back-to-back configuration the length of propagation through the medium can be substantial: If the hard vertex is close to the surface and the near side parton travels outward (as is the case for most triggered hadrons), the away side parton has to cross the whole expanse of the medium, i.e. a length of $\sim 10$ fm before it reaches the opposite surface. At such length- and  timescales, the problem cannot be approximated by a static situation: both longitudinal and transverse flow substantially alter the distribution of matter during the propagation. To illustrate the point more clearly: Naively one would expect that such long pathlengths lead to extremely high energy loss due to the quadratic pathlength dependence of the energy loss on the pathlength $L$. However, due to the accelerated longitudinal and transverse expansion, the medium energy density drops with a high power of time (and hence $L$), more than cancelling the quadratic pathlength dependence. Thus, in a dynamic picture penetration of the fireball core region is possible though rare.

In the following, we present the formalism used to calculate the expected high $p_T$ angular correlation pattern in the presence of an evolving medium and compare with the experimental results.

\section{Partonic energy loss}

We aim to describe data involving trigger hadrons beyond $p_T= $ 8 GeV and associate hadrons with $p_T > 4$ GeV. We assume that at such energies the observed 2-particle correlations are dominated by three contributions: 1) an away side parton emerging from the medium with a large fraction of its original energy left undergoing fragmentation ('punchthrough') 2) a subleading hadron created in the fragmentation of the trigger hard parton ('near side associate') and 3) a subleading hadron created in the fragmentation of the away side parton ('away side associate', only for punchthrough).). We posit that the effect of the medium can completely be treated on the partonic level by inducing partonic energy loss but does not substantially influence the fragmentation process. 

We note that any parton fulfilling the trigger and leading to associated yield must have an energy in excess of 12 GeV (typically 15-20 GeV). Making a simple estimate for the length scale for fragmentation as $L_{\rm frag} = c \cdot E_{\rm part}/Q^2_{\rm had}$ with $Q_{\rm had} \approx 1$ GeV we find length scales of $\sim4$ fm which is enough to carry most partons out of the dense regions of the fireball, thus the assumption seems not unreasonable. 

On the other hand, based on the low $p_T$ correlation pattern and our results in \cite{Mach} we argue that the dominant part of the energy lost to the medium does not appear as an accompanying cone but rather excites collective modes of the medium and hence is observed in the momentum region below $1.5-2$ GeV where the medium shows still collevtivity  (indicated by the validity of a hydrodynamical description for the $p_T$ spectra, cf. e.g. \cite{Kari-Hydro}). Thus, we do not expect the 'lost' energy to contribute substantially to high $p_T$ yield in addition to the sources 1) to 3) listed above.

For the time being we focus on central collisions only. 
Key quantity for the calculation of jet energy loss is the local transport coefficient
$\hat{q}(\eta_s, r, \tau)$ which characterizes the squared average momentum transfer
from the medium to the hard parton per unit pathlength. Since we consider a time-dependent
inhomogeneous medium, this quantity depends on spacetime rapidity $\eta_s = \frac{1}{2}\ln \frac{t+z}{t-z}$,
radius $r$ and proper time $\tau = \sqrt{t^2-z^2}$ (for the time being we focus on central collisions only).
The transport coefficient is related to the energy density of the medium as
$\hat{q} = c \epsilon^{3/4}$ with $c=2$ for an ideal quark-gluon plasma (QGP) \cite{Baier}. In the present study, we have adjusted $c$ such that the observed nuclear suppression is reproduced, resulting in $c=4$.

In order to find the probability for a hard parton $P(\Delta E)$ to lose the energy
$\Delta E$ while traversing the medium, we make use of a scaling law \cite{JetScaling} which allows
to relate the dynamical scenario a static equivalent one by calculating the following quantities
averaged over the jet trajectory $\xi(\tau):$

\begin{equation}
\label{E-omega}
\omega_c({\bf r_0}, \phi) = \int_0^\infty d \xi \xi \hat{q}(\xi)
\end{equation}
and
\begin{equation}
\label{E-qL}
\langle\hat{q}L\rangle ({\bf r_0}, \phi) = \int_0^\infty d \xi \hat{q}(\xi)
\end{equation}

as a function of the jet production vertex ${\bf r_0}$ and its angular orientation $\phi$. We set $\hat{q} \equiv 0$ whenever the decoupling temperature of the medium $T = T_F$ is reached.
In the presence of flow, we follow the prescription outlined in \cite{Jet-Flow,Urs2} and replace

\begin{equation}
\label{E-qhat}
\hat{q} = c \epsilon^{3/4}(p) \rightarrow c \epsilon^{3/4} (T^{n_\perp  n_\perp})
\end{equation}

with

\begin{equation}
\label{E-jetflow}
T^{n_\perp n_\perp} = p(\epsilon) + \left[ \epsilon + p(\epsilon)\right] \frac{\beta_\perp^2}{1-\beta_\perp^2}
\end{equation}

where $\beta_\perp$ is the spatial component of the flow field orthogonal to the parton trajectory.
In the above two expressions, the spacetime dependence $(\eta_s,r,\tau$) of pressure $p$ and energy-momentum tensor
$T^{n_\perp n_\perp}$ have been suppressed for clarity.

Using the results of \cite{QuenchingWeights}, we obtain $P(\Delta E)$ from $\omega_c$ and $\langle\hat{q}L\rangle$
as a function of jet production vertex and the angle $\phi$ from

\begin{widetext}
\begin{equation}
P(\Delta E) = \sum_{n=0}^\infty \frac{1}{n!} \left[ \prod_{i=1}^n \int d \omega_i \frac{dI(\omega_i)}{d \omega}\right]
\delta\left( \Delta E - \sum_{i=1}^n \omega_i\right) \exp\left[-\int d\omega\frac{dI}{d\omega} \right]
\end{equation}

\end{widetext}

which makes use of the distribution $\omega \frac{dI}{d\omega}$ of gluons emitted into the jet cone.
The explicit expression of this quantity for the case of multiple soft scattering can be found in \cite{QuenchingWeights}.

The medium enters via the local energy density $\epsilon(\eta_s,r,\tau)$ in Eq.~(\ref{E-qhat}) and flow $\beta_\perp$ transverse to the jet propagation direction in Eq.~(\ref{E-jetflow}), and it is immediately obvious from Eqs.~(\ref{E-omega}, \ref{E-qhat}) that a strongly dropping energy density as a function of space or time will cancel the quadratic pathlength dependence of e.g. $\omega_c$ in the case of a constant $\hat{q}$.

To describe the medium we use a parametrization of the spacetime evolution which is based on a simultaneous fit to hadronic spectra and Hanbury-Brown-Twiss correlation radii at low $p_T$ \cite{RenkSpectraHBT} (see also \cite{Synopsis} for details of the formalism) and which has been used to successfully predict the emission of thermal photons \cite{Photons_RHIC1, Photons_RHIC2}. We have used the same description to study the excitation of Mach cones by high $p_T$ jets \cite{Mach} and the sensitivity of the nuclear modification factor  
\begin{equation}
R_{AA}(p_T,y) = \frac{d^2N^{AA}/dp_Tdy}{T_{AA}(0) d^2 \sigma^{NN}/dp_Tdy}
\end{equation}
to assumptions about flow in the evolution dynamics \cite{Jet-Flow}. We therefore believe that the model gives a fair representation of the relevant physics of bulk matter in both the early and late stages of the evolution and incorporates the correct amount of quenching demanded by $R_{AA}$.

\section{Monte Carlo Sampling of the trigger conditions}

The approach chosen here is similar to \cite{Dainese} in the sense that we perform a Monte Carlo (MC) sampling of the trigger conditions; with the crucial difference that the model presented here includes the full transverse and longitudinal expansion dynamics.

For central collisions, we assume without loss of generality that the near side parton always propagates into the negative $x$ direction in the transverse $(x,y)$ plane. If we define the $-x$ direction as $\pi$, the away side parton then propagates almost at angle zero, modulo the intrinsic $k_T$ of order $\sim 1$ GeV (the resulting angular spread is small given that we're mainly interested in partons of 15 GeV and above).  

We start by creating jet vertices based on the nuclear overlap function $T_{AA}({\bf b}) = \int dz \rho^2({\bf b},z)$. Next we specify the jet energy $E_{\rm jet}$ by randomly sampling the pQCD parton spectra above the trigger energy.

Basis for the sampling is the leading order (LO) pQCD expression for the back-to-back production of two partons in the transverse plane
\begin{equation}
\label{pQCD}
\frac{d\sigma^{AB \rightarrow kl+X}}{dp_T^2dy_1dy_2}  =
\sum_{ij}  x_1 f_{i/A}(x_1)x_2 f_{j/B}(x_2) 
\frac{d\hat{\sigma}}{dt}^{ij\rightarrow kl} \negthickspace \negthickspace \negthickspace 
\negthickspace \negthickspace \negthickspace \negthickspace \negthickspace 
(\hat{s},\hat{t},\hat{u}).
\end{equation}

where $f_{i(j)/A(B)}(x_{i(j)}, Q^2)$ denote the nuclear parton distributions \cite{NPDF} (in which we have suppressed the scale dependence on $Q^2$ for clarity in the above expression) and $\frac{d\hat{\sigma}}{dt}^{ij\rightarrow kl}
(\hat{s},\hat{t},\hat{u})$ are the leading order pQCD subprocesses for all different combination of parton types $i,j,k,l$ (explicit expressions can be found in \cite{pQCD-Xsec}). The fractional momenta of the colliding partons $i$ and $j$ are given by 

\begin{equation}
x_{1,2} = \frac{p_T}{\sqrt{s}} \left[\exp(\pm y_1) + \exp(\pm y_2)\right]
\end{equation}.

In \cite{Kari2} it has been shown that this framework, when folded with the fragmentation functions and supplemented with a $K-$factor to adjust overall normalization, is able to describe inclusive production of charged hadrons in p-p and p-A collisions for a wide range of energies. This indicates that NLO effects can largely be absorbed into the $K-$factor. Note that such a constant normalization factor drops out in ratios such as $R_{AA}$ or the yield per trigger of dijets we're interested in, thus LO pQCD can be expected to work reasonably well. However, we do introduce intrinsic $k_T$ to account for the correct geometry and the broadening of the observed away side jet (see below).

From Eq.~(\ref{pQCD}) we also determine the probability that the parton is a gluon (by setting e.g. $k=g$). We propagate the near side parton to the surface and determine $\omega_c$ and $\langle\hat{q}L\rangle$ by evaluating Eqs.~(\ref{E-omega}, \ref{E-qL}) along the path. The resulting values serve as input for the probability distribution of energy loss $P(\Delta E)$ as determined in \cite{QuenchingWeights}.
Note that in addition to a continuous distribution $P(\Delta E)$ there is also a discrete part $P(0)$ which reflects the probability to loose no energy. This part is finite (but often small) for all possible vertices and paths.

Often the plasma frequency $\omega_c$ is far above the available jet energy and $P(\Delta E)$ thus extends to energies $\Delta E \gg E_{\rm jet}$. This reflexts the fact that the radiative energy loss in \cite{QuenchingWeights} is derived in the limit of infinite parton energy and translates into an uncertainty. Two different prescriptions have been considered as the upper and lower limit quantifying this uncertainty:

\begin{itemize}
\item Reweighting: Here, one truncates $P(\Delta E)$ at $\Delta E = E_{\rm jet}$ and renormalize to unity by dividing out the factor $\int_0^E P(\Delta E)$.
\item No-reweighting: Alternatively, one may truncate $P(\Delta E)$ at $\Delta E = E_{\rm jet}$ and consider the parton energy to be zero if $\Delta E$ exceeds $E_{\rm jet}$.
\end{itemize}

The quenching power of the medium is higher in the non-reweighted case. However, note that reweighting is only a reasonable prescription if $E_{jet}$ is above the $\Delta E$ at which $P(\Delta E)$ peaks. If that is not the case (as regularly observed for away side partons propagating through dense regions in the simulation), reweighting becomes meaningless as easily seen from the fact that under some conditions the average energy loss after reweighting may decrease if $\hat{q}$ is increased. On the other hand, we are not interested in details of energy losses of order $O(E_{jet})$ - once the energy loss reaches a sizeable fraction of the parton energy, the event will fall below the trigger/associate threshold and the actual $\Delta E$ is irrelevant. Note that this is very different from a calculation of $R_{AA}$ where the partons experiencing substantial energy loss do not fall below a threshold but reappear elsewhere in the spectrum. The dijet measurement is only sensitive to the probability to have little or no energy loss, and this should be independent of reweighting considerations as $\Delta E$ is always a small fraction of $E_{jet}$ for almost all hadrons above the trigger or associate cut. Thus, we do not apply the reweighting in the following.

We determine the actual energy loss of the near side parton by sampling $P(\Delta E)$. To find the energy of the leading hadron, we use the KKP fragmentation functions \cite{KKP} to find the probability $D_i(z, \mu)$ for a parton $i$ to fragment into a hadron carrying the momentum fraction $z$ at a typical hadronic scale $\mu$. Thus, we evaluate the schematical expression

\begin{equation}
d\sigma_{med}^{AA\rightarrow h+X} = \sum_f d\sigma_{vac}^{AA \rightarrow f +X} \otimes P_f(\Delta E) \otimes
D_{f \rightarrow h}^{vac}(z, \mu_F^2)
\end{equation} 

in a MC framework where $d\sigma_{vac}^{AA \rightarrow f +X}$ is the pQCD expression for the hard scattering process.

If the resulting hadronic $P_{\rm had} = z p_{\rm jet} \approx z E_{\rm jet}$ fulfills the trigger condition we accept the event and proceed with the calculation of associated hadrons and the away side parton, otherwise we reject the event and continue the MC sampling by generating a new vertex.

\section{Monte Carlo Sampling of the away side and associated hadrons}

If an event fulfillig the trigger has been created, we determine the intrinsic $k_T$ being added to the away side parton momentum. We sample a Gaussian distribution chosen such that the widening of the away side cone without a medium is reproduced. Since this is a number of order $1$ GeV whereas partons fulfilling trigger conditions have frequently in excess of $15$ GeV we note that this is a small correction.

We treat the far side parton exactly like the near side parton, i.e. we evaluate Eqs.~(\ref{E-omega}, \ref{E-qL}) along the path and find the actual energy loss from $P(\Delta E)$ with $\omega_c, \langle\hat{q}L\rangle$ as input. If the away side parton emerges with a finite energy, we use the fragmentation function $D_i(z, \mu)$ to determine the momentum of the away side hadron. If this momentum fulfills the cut for associated particle production, we count the event as 'punchthrough'.

In addition, we allow for the possibility that the fragmentation of near and away side parton produces more than one hard hadron. Since we are predominantly interested in the quenching properties of the medium, we sample this probability using the measured probability distribution $A_i(z_F)$ of associated hadron production in d-Au collisions \cite{Dijets, Dijets2} as a function of $z_T$ where $z_T$ is the fraction of the trigger hadron momentum carried by the associated hadron. We include a factor $\theta(E_{\rm jet} - E_{\rm trigger} - \Delta E - E_{\rm assoc})$ on the near side and $\theta(E_{\rm jet} - E_{\rm punch} - \Delta E - E_{\rm assoc})$ on the far side to make sure that energy is conserved. Note that associated production on the far side above the $p_T$ cut is only possible if a punchthrough occurs. We count these events as 'near side associate production' and 'away side associated production'.

Both the fragmentation function $D_i(z, \mu)$ and the associated production probability distribution $A_i(z_F)$ are steeply falling functions of $z, z_F$. They have to be regulated by a lower cutoff, otherwise the fragmentation function will be evaluated in a regime where it is not applicable (In calculations starting from the hadron spectrum, a lower cutoff at the hadronic scale $P_T$, i.e. $z_{min}(P_T) = \frac{2 P_T}{\sqrt{s}}$ arises from  the fact that no parton can exceed the $p_T$ of the kinematic limit. In the present Monte Carlo model where the parton spectrum is determined first it is difficult to implement the same hadronic $P_T$ dependence of the cutoff, hence we follow a more heuristic approach.).

We choose the cutoff in such a way that the measured yield per trigger as a function of the cut on associate particle production is reproduced in the case of d-Au collisions, i.e. without energy loss. We find that  $z^{\rm min} = 0.05$ and $z_F^{\rm min} = 0.1$ gives a fair representation of the data and use the same values also in the presence of a medium. The resulting difference then only reflects the effect of the medium on the parton energies.

Thus, the yield per trigger on the near side is determined by the sum of all 'near side associate production' events divided by the number of events fulfilling the trigger, the yield per trigger on the away side is given by the sum of 'punchthrough' and 'away side associated production' divided by the number of events. These quantities can be directly compared to experiment.

\section{Results}

\begin{figure*}[htb]
\epsfig{file=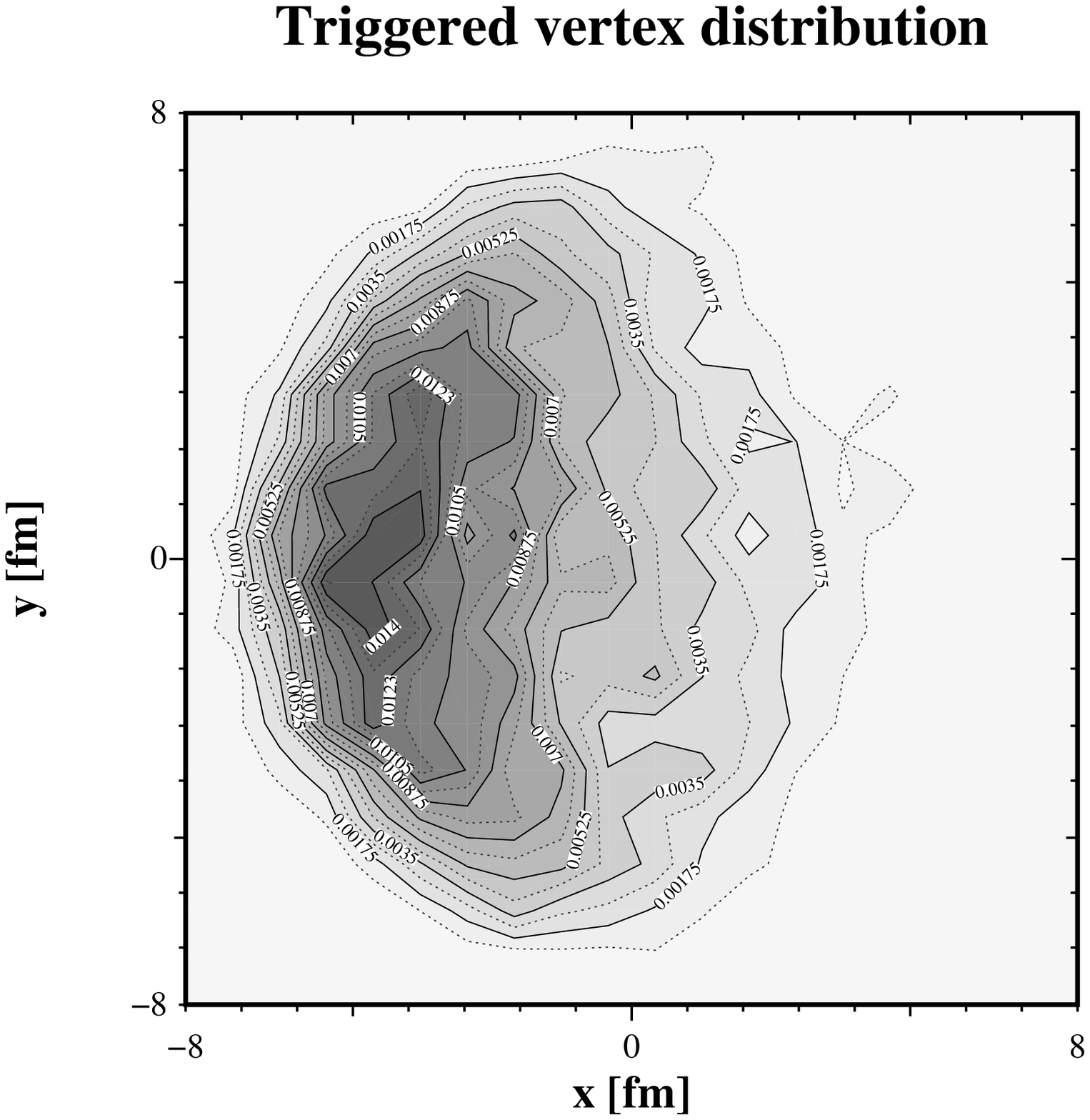, width=7.5cm}\epsfig{file=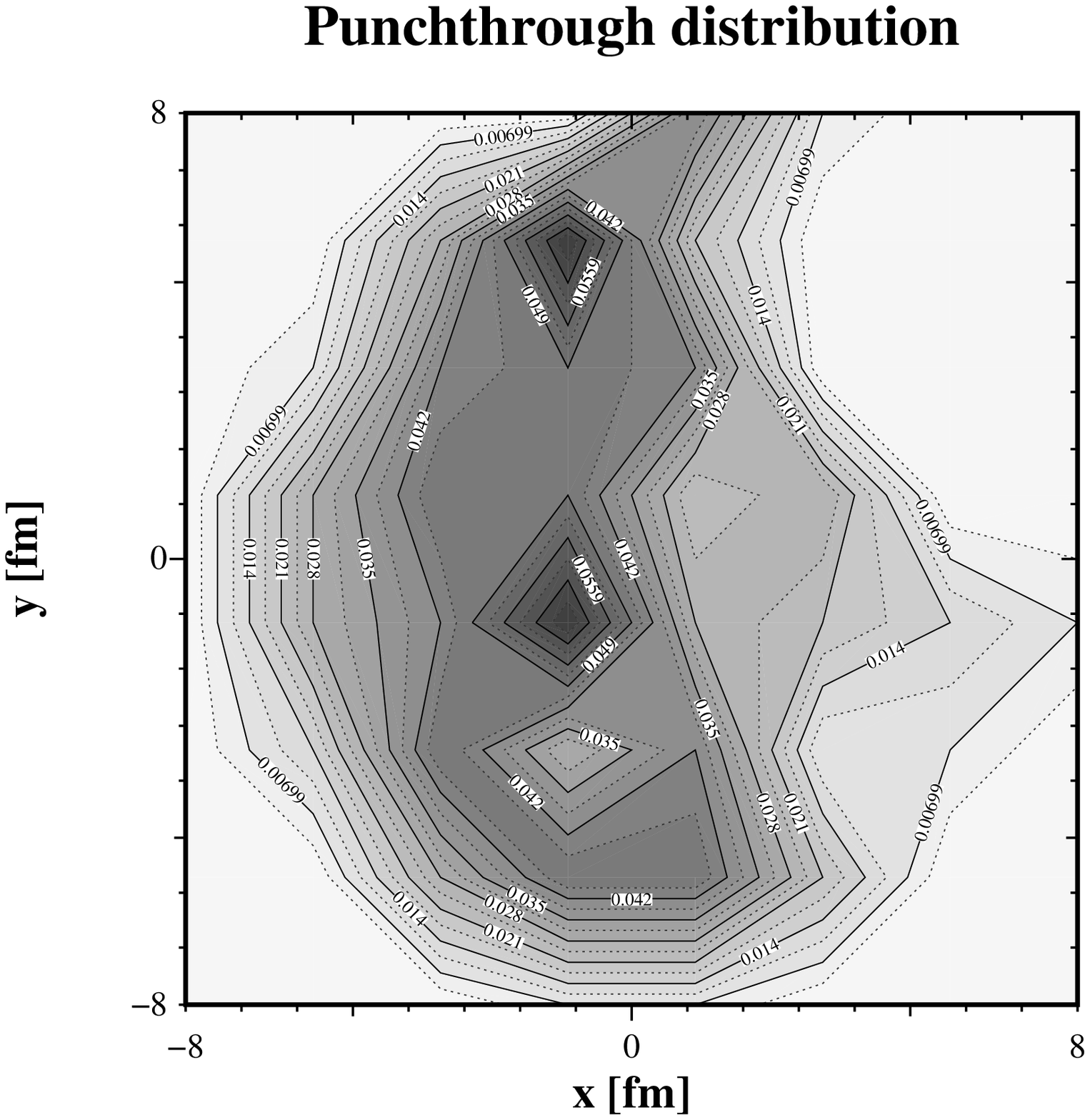, width=7.5cm}
\caption{\label{F-vertex}Left panel: Normalized density of vertices for events with a trigger hadron above 8 GeV. Right panel: Normalized density of vertices for which an associate hadron in the momentum region 4 $< p_T <$  GeV is found. The densities are based on 8600 simulated events, in all cases the near side parton propagates into negative x direction.}
\end{figure*}

In Fig.~\ref{F-vertex} we show the resulting distribution of vertices leading to a valid event and the position of vertices for which a dijet was observed. We can infer a few interesting observation from the distribution:

First, most triggered events cluster around the near side surface within a 'skin' of about 3 fm thickness. This is expected, as this is the region in which the energy loss of the near side parton is expected to be small. However, a sizeable number of triggered events also originates from the deep central region. This 'skin' effect has already been discussed in e.g.  \cite{Dainese}. 

More interesting is the distribution of events leading to an associated away side hadron ('punchthrough'). This distribution is manifestly different from the vertex distribution for triggered events. First, its main strength comes from the region around $x\sim 0$, hence events with equal pathlengths on the near and away side are more likely to result in dijets. Second, it extends much further into the large $R$ region (the 'halo') than the distribution of triggered vertices.  In the halo (i.e. in the tail of the nuclear overlap distribution the energy density is low to start with, and by the time a parton from the halo propagates into the center the energy density there has been reduced sizeably by the fireball expansion, hence such a parton never encounteres significant energy densities and dijet events are hence favoured in this region. In addition, there may be some tendency for dijet events to accumulate in the surface region close to $x=0,y\sim \pm6$ fm where the pathlength through matter is small for both near and away side parton. Note that while both plots are in principle expected to be symmetrical around the x-axis, in practice they are not due to the finite size of the sample of simulated events.

Let us briefly remark on the mechanism which underlies this distribution: We observe that about 75\% of all trigger partons probe the discrete part of the quenching weight, i.e. did not undergo any energy loss. This is in agreement with the observation made in \cite{Dainese} that the energy loss probability of observed hadrons is dominated by the discrete part. However, the crucial difference is that the expansion dynamics and subsequent drop in density allows the region in which the discrete quenching weight is still relevant to move further into the fireball center.

This in turn has a strong influence on the pathlength distribution of the away side parton --- if the production vertex is not confined to the surface, the average pathlength of the away side parton is significantly reduced (and indeed the distribution favours about equal pathlengths for near and away side parton). The magnitude of energy loss on the away side for observed hadrons is rather strongly momentum dependent: While only $\sim 30$\% of all hadrons in the 4-6 GeV momentum bin probe the discrete quenching weight in the simulation, the fraction increases to $\sim60\%$ in the 6+ GeV associate hadron momentum region. Thus, the larger the difference between trigger energy and associate hadron cut, the more is the sensitivity to the continuous part of the energy loss probability increased.

\begin{figure*}[htb]
\epsfig{file=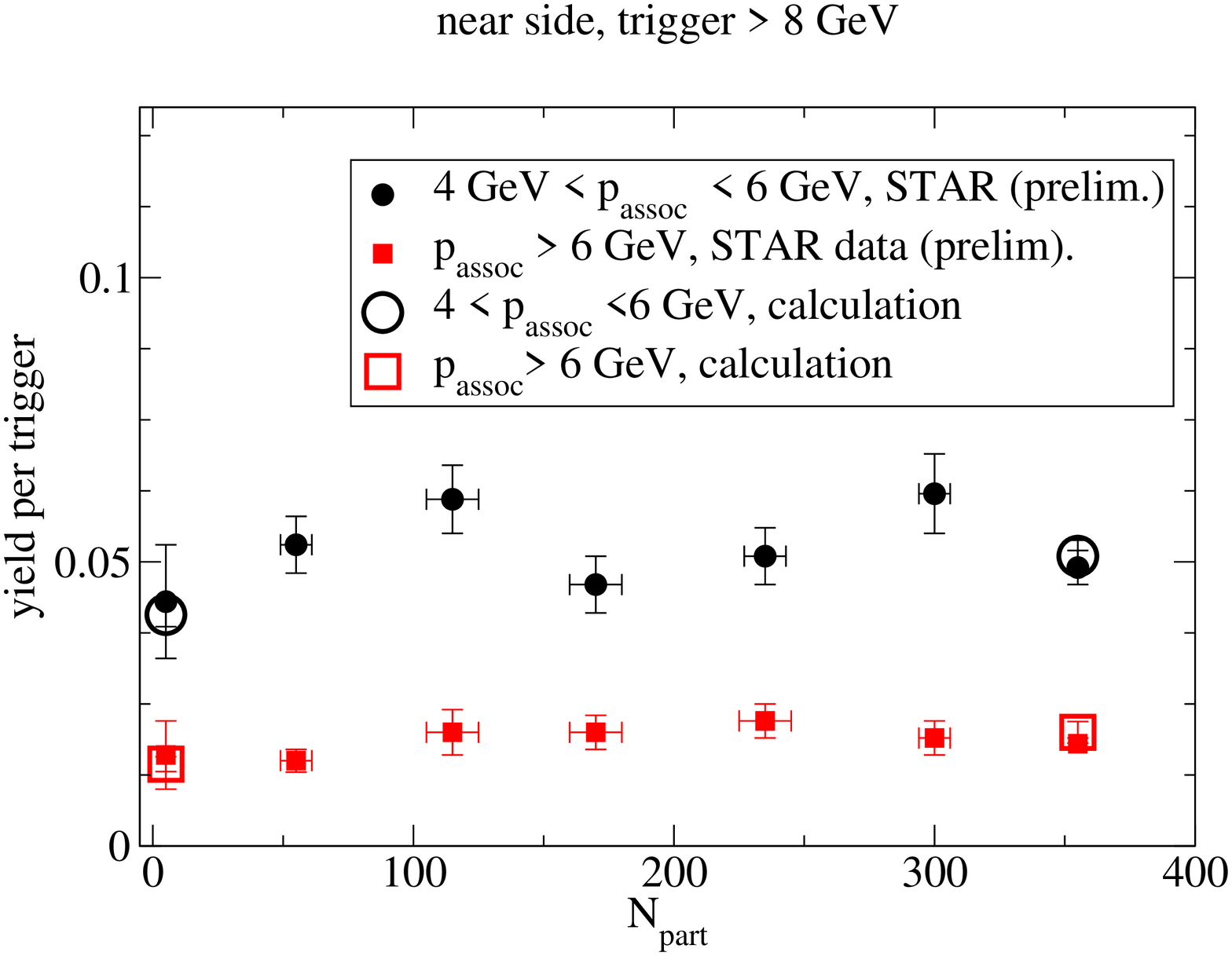, width=7.5cm}\epsfig{file=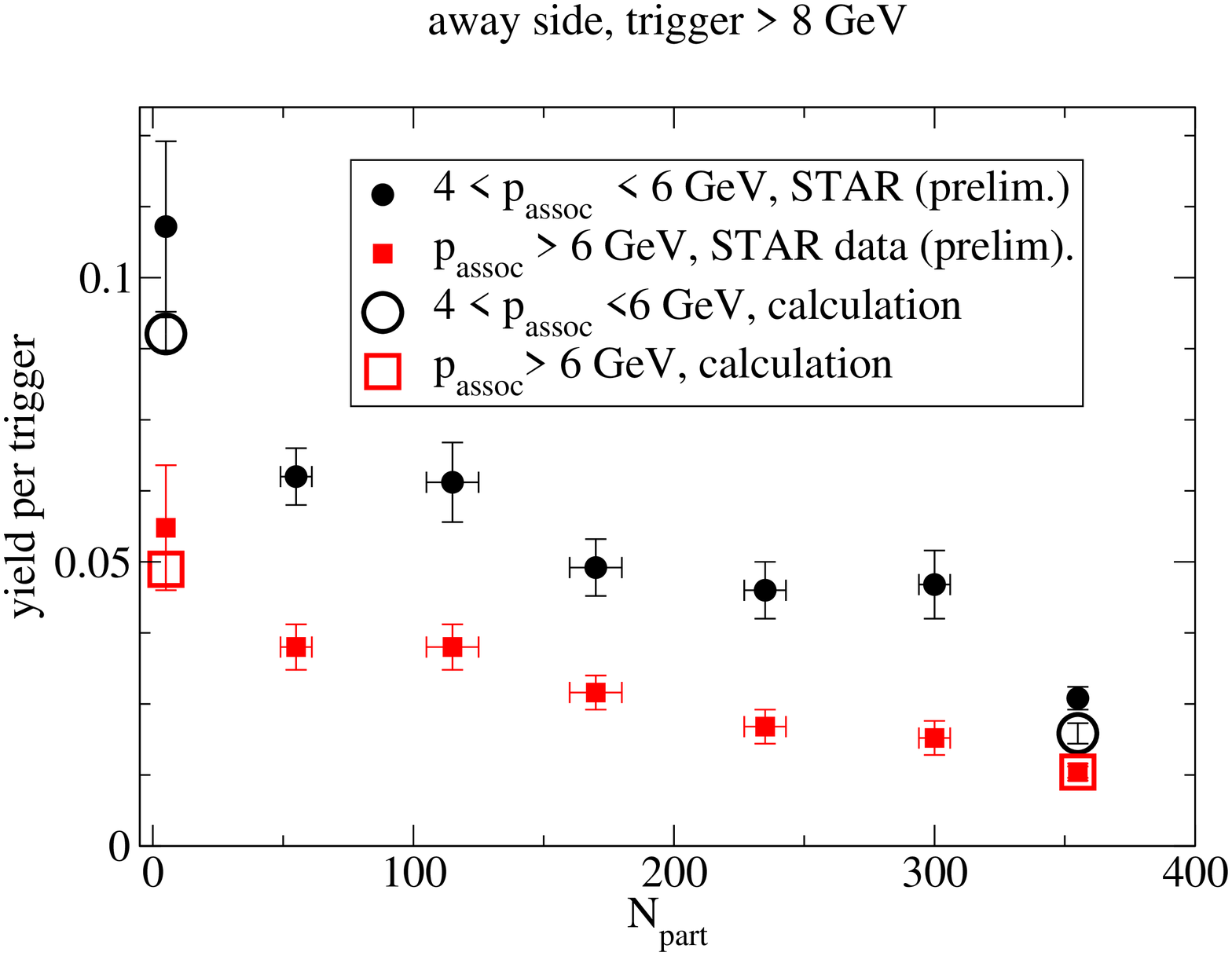, width=7.5cm}
\caption{\label{F-yield}Yield per trigger on the near side (left panel) and away side (right panel) for trigger hadron above 8 GeV in the model calculation as compared with STAR data.}
\end{figure*}

We show the resulting yield per trigger compared with the STAR data in Fig.~\ref{F-yield} (since the model calculation has only been performed for central collisions and vanishing medium, we cannot address the full range of $N_{part}$ probed by the experiment). The results of the model calculation agree surprisingly well with the data, considering that all hadronization parameters have been fixed without reference to the medium and that all quenching properties of the medium are completely determined by the description of $R_{AA}$ \cite{Jet-Flow}.

It is reassuring that the model works especially well for the $p_T > 6$ GeV associate hadron binning, as in this region the assumption of fragmentation being the only relevant mechanism for hadronization is better justified. It is plausible that the discrepancy of $\sim 30\%$ in the 4-6 GeV window on the away side can be traced back to a recombination contribution (not contained in the present model) as the hadron-species averaged contribution of recombination to the hadron spectrum in central Au-Au collisions bewteen 4 and 6 GeV is of the same order of magnitude (cf. Fig. 4 in \cite{Reco}).

\section{Conclusions}

We have evaluated radiative energy loss for high $p_T$ partons within a dynamical model for the evolution of hot and dense matter. The same energy loss formalism which describes $R_{AA}$ is in such a model also able to describe the correct amount of experimentally observed high $p_T$ triggered high $p_T$ associated hadron ('dijet') events under the assumption that these events represent hard partons emerging from the medium and fragmenting in vacuum. 

Note that in this picture, the central region is still on average very opaque. The average value of the energy loss for a quark propagating outward from the center of the fireball is about  $\langle \Delta E \rangle = 23$ GeV -  on average partons from the center cannot reach the surface. Thus, most of the observed trigger and dihadron correlation yield coming from the central region of the medium represents a relatively rare class of events in which the low $\Delta E$ tail of the probability distribution or the discrete quenching weight is probed. 

We have argued based on the distribution of production vertices that a sizeable amount of observed associated hard hadrons propagates a sizeable distance $> 5$ GeV through the medium. For such long timescales, the drop in energy density from both longitudinal and transverse expansion cannot be neglected, therefore a dynamical model evolution is crucial. Especially events leading from the halo region which never encounter a significant energy density do not emerge in a static description of jet quenching. Another crucial ingredient of the simulation is the discrete part of the quenching weight, i.e. the probability to have no energy loss in spite of a sizeable pathlength through the medium. The magnitude of this probability likewise depends on the line-integral over the dropping medium density and is thus also on average significantly larger in a dynamical picture than in a static model.  While this probability is still numerically small, in the calculation of dijet yields it competes with other small quantities which allow for hard dihadron production like the probability to select a parton from the high $p_T$ tail of the parton spectrum or to fragment into a hadron with $z\approx 1$, probing the tail of the fragmentation function or to select a vertex at large radius where $T_{AA}$ is small. Since the dihadron yield itself is not numerically large, all these different effects contribute in a way that is characteristic for the medium evolution.

We conclude that radiative energy loss is well able to account for both $R_{AA}$ and the observed pattern of dijet events. While the central part of the medium is very dense, the model simulation indicates that it is by no means completely black, and high $p_T$ triggered events seem indeed to probe the densest part of the medium to some degree. 

Recently, elastic energy loss has been suggested as an additional ingredient to describe the nuclear suppression (see e.g. \cite{Elastic}). To the degree that the magnitude of  this contribution is parametrically given by $\Delta E \sim \int_0^\infty d \xi \hat{q}(\xi)$ (note the absence of a factor $\xi$ as compared with Eq.~(\ref{E-omega}) for radiative energy loss), the results of the present investigation place tight limits on the importance of collisional energy loss: Since the yield per trigger of hard dihadrons is determined by a delicate interplay between differences in average near and away side pathlength, in turn leading to different energy loss probabilities, the scenario cannot easily accomodate a different pathlength dependence. In particular, the collisional integral above  is dominated by early times when the density is largest (whereas the radiative integral Eq.~(\ref{E-omega}) gets most support around 3-4 fm/c) and is completely insensitive to late times (and consequently long paths). However, this implies that there would be no additional suppression for the away side partons in dihadron measurements as compared to d-Au; while the total yield would go down like $R_{AA}$, the yield per trigger (which in the present simulation rests on the systematic difference in pathlength for near and away side parton and its weighting via Eq.~(\ref{E-omega})) would be unchanged. A more detailed future investigation will make this point more quantitative.

\begin{acknowledgments}

I would like to thank V.~Ruuskanen, K.~Eskola, B.~M\"{u}ller, S.~Bass, R.~Fries,  P.~Jacobs J.~Rak and J.~Ruppert for valuable comments and discussions. In addition, support from the NERSC computing center is gratefully acknowledged. This work was financially supported by the Academy of Finland, Project 206024. 

\end{acknowledgments}

\end{document}